\documentclass[useAMS,usenatbib]{mn2e} 
\usepackage{graphicx} 
\usepackage{amssymb}
\newcommand{\mnras}{MNRAS}
\newcommand{\aj}{ApJ}
\newcommand{\apj}{ApJ}
\newcommand{\apjs}{ApJ}
\newcommand{\apjl}{ApJ}
\newcommand{\aap}{A\&A}
\newcommand{\apss}{ASS}
\newcommand{\araa}{ARA\&A}
\newcommand{\memsai}{Memorie della Societa Astronomica Italiana}

\newcommand{\pasp}{PASP}
\newcommand{\nat}{Nature}
\newcommand{\icarus}{Icarus}

\newcommand{\gppr}{\stackrel{>}{\scriptstyle \sim}}
\newcommand{\gappr}{\raisebox{-0.4ex}{$\gppr$}}

\newcommand{\Msun}{\mbox{$\mathrm{M}_{\odot}$}}

\newcommand{\Porb}{\mbox{$P_{\mathrm{orb}}$}}

\newcommand{\Ion}[2]{#1{\,\scriptsize #2}}

\title[M dwarf  magnetic activity]{M dwarf companions  to white dwarfs
  I: relating magnetic activity, rotation and age}
\author[A. Rebassa-Mansergas et al.]{A. Rebassa-Mansergas$^1$, M.R. Schreiber$^{1,2}$
B.  T.  G\"ansicke$^3$\\
$^{1}$ Departamento de F\'\i sica y Astronom\'\i a, Universidad de Valpara\'\i so, 
Avenida Gran Bretana 1111, Valpara\'\i so, Chile \\
$^{2}$ Millenium Nucleus "Protoplanetary Disks in ALMA Early Science,"
Universidad de Valparaiso, Chile\\
$^{3}$ Department of Physics, University of Warwick, Coventry CV4 7AL, UK \\
}

\begin{document}
\date{Accepted 2012. Received 2012; in original form 2012}
\pagerange{\pageref{firstpage}--\pageref{lastpage}} \pubyear{2012}
\maketitle

\begin{abstract}
We  make use  of  the largest  and  most homogeneous  sample of  white
dwarf/M dwarf (WD/dM) binaries from the Sloan Digital Sky Survey (SDSS
DR7)  to investigate  relations between  magnetic  activity, rotation,
magnetic  braking and  age in  M  stars. These  relations are  studied
separately for  close WD/dM binaries that underwent  a common envelope
phase and  thus contain  tidally locked and  hence rapidly  rotating M
dwarfs, and  for wide  WD/dM binaries that  never interacted.  For the
wide WD/dM binary  sample we find that the  M dwarf activity fractions
are  significantly higher  than those  measured in  single M  stars of
spectral type M0 to M5.  We  attribute this effect as a consequence of
M dwarfs in wide SDSS  WD/dM binaries being, on average, significantly
younger and hence more active  than the field M dwarf population.  The
measured M  dwarf activity  fractions in wide  WD/dM binaries  show as
well a significant increase from  spectral types M3 to M5, where these
low-mass  stars  become fully  convective.   This provides  additional
observational evidence for magnetic braking being less efficient below
the  fully convective boundary,  in agreement  with the  hypothesis of
fully convective  stars having considerably  longer activity lifetimes
than  partially convective  stars.   The  M dwarfs  in  all our  close
binaries are active, independently  of the spectral type, giving robust
observational  evidence for  magnetic activity  being enhanced  due to
fast rotation. The rotational velocities  of the M dwarfs in our close
binary sample are significantly higher than seen among field M dwarfs,
however  the  strength  of  magnetic  activity  remains  saturated  at
$\log$\,L$_\mathrm{H\alpha}$/L$_\mathrm{bol}    \sim    -3.5$.    This
unambiguously confirms  the M dwarf  saturation-type rotation-activity
relation.
\end{abstract}


\label{firstpage}

\section{Introduction}

The mechanisms that induce magnetic  activity in field stars have been
intensively  studied and  discussed  over the  last  decades and  this
effort is now converging in a fairly consistent scenario that explains
both magnetic field generation  and angular momentum evolution in both
partially and fully convective stars.

Magnetic fields in partially convective  (F,G,K and early M) stars are
believed  to  be  generated  at  the  transition  region  between  the
radiative  interior and the  differentially rotating  outer convective
zone (tachocline) through a  solar-type $\alpha\Omega$ dynamo. In this
scenario magnetic fields are generated as a result of a combination of
differential  rotation  ($\Omega$   effect)  and  cyclonic  convection
($\alpha$  effect) \citep{parker55-1,  leighton69-1, spiegel+zahn92-1,
  charbonneau05-1,  browning08-1}.  Lower  mass M  dwarfs  of spectral
type later  than $\sim$M3  are fully convective  and therefore  do not
posses a tachocline, however  display clear signs of magnetic activity
\citep[e.g.][]{delfosseetal98-1}. Such  stars are suggested  to rotate
as   rigid   bodies    \citep{barnesetal05-1},   implying   that   the
$\alpha\Omega$  dynamo cannot  be responsible  for  producing magnetic
fields.  Instead, an $\alpha^2$  dynamo has been proposed for magnetic
field  generation in  fully  convective stars  \citep{raedleretal90-1,
  chabrier+kueker06-1}.

The efficiency  of both the  $\alpha\Omega$ dynamo and  the $\alpha^2$
dynamo are expected  to be strongly correlated with  the Rossby number
\citep[$R_{0}  =  P_\mathrm{rot}/\tau_0$,  with  $P_\mathrm{rot}$  the
  rotational   period    and   $\tau_0$   the    convective   overturn
  timescale]{noyesetal84-1}.   In partially convective  stars magnetic
activity  rises for  decreasing Rossby  numbers, implying  that larger
rotational  velocities  lead to  higher  levels  of magnetic  activity
\citep[for  a  given  $\tau_0$,][]{hartmann+noyes87-1}.  Similarly,  a
correlation between  activity and rotation  is also expected  in fully
convective  stars in  which magnetic  fields may  be generated  by the
$\alpha^2$   dynamo  \citep{durney+stenflo72-1,  chabrier+kueker06-1}.
Several  early  observational  studies  have explored  the  connection
between magnetic activity  and rotation \citep{wilson66-1, kraft67-1}.
More  recently,   \citet{pizzolatoetal03-1}  showed  that   the  X-ray
emission  increases  with  rotation  for  slowly  rotating  stars  but
saturates below a threshold of  $R_{0} \simeq 0.1$.  The same relation
seems     to    apply    for     the    magnetic     field    strength
\citep{reinersetal09-1}.  Unfortunately,  the phenomenon of saturation
is  not  fully  understood  yet,  one  possible  explanation  is  that
saturation   occurs  due   to   a  change   of  dynamo   configuration
\citep{vilhu84-1, wrightetal11-1}.  An additional explanation  is that
saturation is reached when a certain maximum fraction of the available
energy  flux  (from  convection)  is converted  into  magnetic  energy
\citep{christensenetal09-1}.   For ultracool  dwarfs of  spectral type
M7-9.5 this clear rotation-activity  relation, however, seems to be no
longer valid \citep{mohantyetal02-1, reiners+basri10-1}.  Moreover, it
has  to be  stressed that  rotation  and magnetic  activity in  single
(late-type)    M     dwarfs    might    not     always    be    linked
\citep{west+basri09-1}.

Rotation and magnetic  activity decrease in time \citep{skumanich72-1,
  mestel84-1,     mestel+spruit87-1,    kawaler88-1,    sillsetal00-1,
  westetal08-1} as  a consequence  of magnetic braking,  i.e.  angular
momentum is extracted from the  convective envelope and lost through a
magnetized  wind.   The  braking  timescales strongly  depend  on  the
spectral   type    (or   mass)   of    the   star   \citep{barnes03-1,
  barnes+kim10-1},  which results  in lower-mass  partially convective
stars spinning-down slower than their higher mass counterparts.  Fully
convective  stars  have  even  longer  rotational  braking  timescales
\citep{reiners+basri08-1,     browningetal10-1,    schreiberetal10-1}.
Unfortunately the reason for  magnetic braking being less efficient in
fully  convective stars is  not completely  understood. This  might be
related to  a change  in the magnetic  field topology, which  has been
observed to  switch from less  ordered field structures to  dipoles at
the fully convective boundary going from partially to fully convective
stars    \citep{donatietal08-1,    morinetal08-1,   reiners+basri09-1,
  morinetal10-1}.   However such  a  switch should  lead to  increased
magnetic braking for fully  convective stars, exactly opposite to what
is  indicated by  the observed  rotation  rates.  A  new scenario  for
magnetic     braking     has     been    recently     developed     by
\citet{reiners+mohanty12-1} who  suggest that the  steep transition of
the braking rates  at the fully convective boundary  arises due to the
significant drop in radius, without  invoking any change in the dynamo
theory.

M dwarf companions to white dwarfs that form part of detached binaries
(hereafter  WD/dM  binaries) offer  an  ideal  test  bed for  studying
magnetic  activity, rotation  and  magnetic braking  across the  fully
convective boundary for three  reasons.  (1) WD/dM binaries can easily
be  separated  into  wide  binaries  with  orbital  periods  typically
exceeding $100$\,days, and close  binaries that evolved through common
envelope  evolution  \citep{webbink84-1,zorotovicetal10-1} with  final
orbital periods generally  below one day \citep{nebotetal11-1}.  While
the M dwarfs in wide binaries  should be unaffected by the presence of
the white dwarf, those  in close post-common envelope binaries (PCEBs)
are tidally locked, i.e. they  are rapidly rotating. Comparing the two
samples can provide  clear constraints on the impact  of fast rotation
on activity.   (2) Using PCEBs  with measured orbital periods  one can
furthermore directly relate the strength of activity to the rotational
velocity of  the M dwarf (the  orbital period of the  binary equals to
the rotational period  of the M dwarf which  then gives the rotational
velocity by adopting a spectral  type-radius relation). (3) The age of
the  M  dwarfs in  wide  WD/dM  systems  can be  estimated  relatively
accurately using cooling tracks plus an initial-to-final mass relation
for  their  white dwarf  companions.   This  allows  to test  activity
lifetimes that are presumably related to magnetic braking models.

The first studies  of magnetic activity in M dwarfs  that form part of
WD/dM    binaries    from     the    Sloan    Digital    Sky    Survey
\citep[SDSS,][]{yorketal00-1,   aiharaetal11-1}   were  performed   by
\citet{silvestrietal06-1} and more recently by \citet{morganetal12-1}.
However, these two  studies consider all WD/dM binaries  in SDSS to be
part   of   close   binaries,   which   is  clearly   not   the   case
\citep{schreiberetal10-1,    rebassa-mansergasetal11-1},   and   their
results should be interpreted with extreme caution. Only our dedicated
radial  velocity  survey  of   WD/dM  binaries  from  SDSS  allows  to
accurately  separate wide  binaries (in  which the  stellar components
have  evolved  as  if  they  were single)  from  close  binaries  that
underwent a  common envelope  evolution and therefore  contain tidally
locked    fast    rotating    M    dwarfs    \citep{schreiberetal10-1,
  rebassa-mansergasetal11-1}.   This  is  fundamental to  explore  the
relation  between  activity  and   rotation.   The  largest  and  most
homogeneous  sample  of  WD/dM  binaries from  SDSS  including  radial
velocity       information      has       been       presented      in
\citet{rebassa-mansergasetal12-1}. This sample  forms the input for the
present  paper where  we  analyse  the magnetic  activity  of M  dwarf
components  in  WD/dM  binaries,  thereby testing  possible  relations
between age, activity and rotation.

\section{The SDSS WD/dM binary catalogue}
\label{s-cat}

The  catalogue of  \citet{rebassa-mansergasetal12-1} consists  of 2248
white dwarf-main sequence binaries from  the data release 7 (DR\,7) of
SDSS, of  which 2019  contain an M  dwarf (WD/dM binaries).   The SDSS
WD/dM binary sample  is described in more detail  in this Section.  We
begin introducing  our radial velocity survey that  allows to separate
between  close and  wide binaries.   We  then explain  how SDSS  WD/dM
binary spectra  can be used for obtaining  reliable stellar parameters
of both components and finally give indications of how we estimate the
ages of our systems.

\subsection{Close and wide WD/dM binaries}
\label{s-sample}

The population of WD/dM  binaries contains close binaries (post-common
envelope binaries  or PCEBs) that underwent  common envelope evolution
\citep[][]{webbink84-1,iben+livio93-1,zorotovicetal10-1},   and   wide
binaries that never  interacted. The M dwarfs that  form part of PCEBs
are tidally locked and are therefore fast rotators.  The properties of
the M dwarfs  that are part of wide binaries  should resemble those of
single M dwarfs as the  stellar components in wide WD/dM binaries have
evolved as if they were single stars \citep{willems+kolb04-1}.

To  characterize large  samples of  both  close PCEBs  and wide  WD/dM
binaries we  are performing a large radial  velocity follow-up program
of  SDSS WD/dM binaries.   Our observing  strategy follows  a two-step
procedure. In a first step we  obtain at least two spectra per target,
separated by at least one night,  and identify systems as PCEBs if the
radial  velocities  we  measure  from  these spectra  show  more  than
3$\sigma$ radial velocity variation.   If we do not detect significant
radial  velocity variation,  we consider  the system  as a  wide WD/dM
binary candidate.  This technique  is obviously less sensitive towards
the  detection of  long  orbital period  PCEBs and/or  low-inclination
systems and we predict that, on average, $\sim$16 per cent of the wide
WD/dM   binary    candidates   are   in    fact   unrecognised   PCEBs
\citep{nebotetal11-1}. In  a second step,  the orbital periods  of the
PCEBs identified in  this way are measured from  higher cadence radial
velocity follow-up.

So  far  we have  identified  191 PCEBs  and  1055  wide WD/dM  binary
candidates \citep[e.g.][]{rebassa-mansergasetal11-1} and have measured
the  orbital  periods  of 81  PCEBs  \citep[e.g.][]{schreiberetal08-1,
  rebassa-mansergasetal08-1,                             nebotetal11-1,
  rebassa-mansergasetal12-2}.  These  samples of wide  and close WD/dM
binaries  offer  the  potential  to e.g.   directly  compare  activity
fractions  of slowly  and  fast rotating  M  dwarfs as  a function  of
spectral type.  Furthermore, the rotational velocities of the M dwarfs
in  PCEBs with known  orbital periods  can also  be known  (assuming a
spectral type-radius relation). These  M dwarfs are very fast rotators
and  should populate the  saturated regime,  therefore allowing  us to
test the saturation-type rotation-activity relation.

\subsection{Stellar parameters}
\label{s-param}

\begin{figure}
\includegraphics[angle=-90,width=\columnwidth]{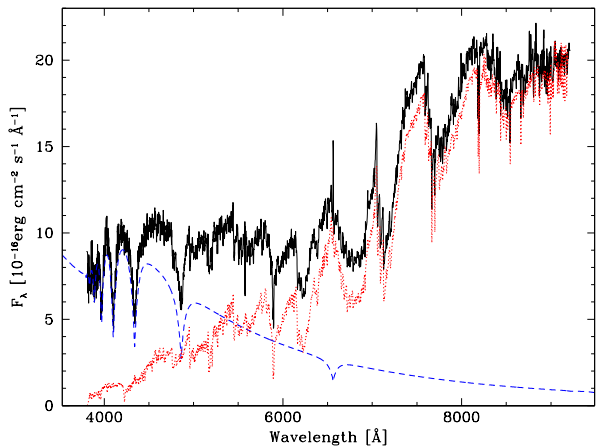}
\caption{\label{f-norm1}  SDSS  spectrum  of  SDSSJ042200.51+073358.9,
  together with the best fit  white dwarf model (blue dashed line) and
  best-fit M dwarf template (red solid dotted line).}
\end{figure}

Reliable stellar  parameters of the  WD/dM binaries are  obtained from
the   SDSS    spectra   using   our    decomposition/fitting   routine
\citep{rebassa-mansergasetal07-1}.   The  SDSS  spectrum is  initially
fitted with  a two-component  model using a  grid of  observed M-dwarf
templates  and   a  grid  of  observed  white   dwarf  templates  (see
Figure\,\ref{f-norm1}),  and  the spectral  type  of  the  M dwarf  is
determined.   Then,  the best-fit  M  dwarf  template,  scaled by  the
appropriate flux scaling factor,  is subtracted and the residual white
dwarf  spectrum  is  fitted with  a  model  grid  of DA  white  dwarfs
\citep{koester10-1} to determine the white dwarf effective temperature
and  surface  gravity.  From  an  empirical spectral  type-radius-mass
relation   for   M-dwarfs   \citep{rebassa-mansergasetal07-1}  and   a
mass-radius         relation          for         white         dwarfs
\citep{bergeronetal95-2,fontaineetal01-1} the masses  and radii of the
M dwarf and the white dwarf components are calculated.

In the context of the present paper the spectral types of the M-dwarfs
are  of fundamental  importance.  We  therefore compared  the spectral
types  of  our  template   fitting  routine  with  those  obtained  by
\citet{westetal11-1}.  Fitting several hundred  M dwarfs from the list
of  \citet{westetal11-1}, equally distributed  in spectral  type, with
our M dwarf  templates and adopting an uncertainty  of half a spectral
type  sub-class,   we  find   that  98  per   cent  of   our  spectral
classifications  agree with those  of \citet{westetal11-1}  within one
spectral subclass.

\subsection{Ages}
\label{s-age}

For a given mass, the magnetic activity of M dwarfs is closely related
to     their     rotational      velocity     (see     for     example
\citealt{cardini+cassatella07-1}).  If other  effects, such as a close
companion, do  not affect the  star, magnetic braking is  slowing down
the  rotation with time.   The magnetic  activity in  our sample  of M
dwarfs that  form part  of wide SDSS  WD/dM binaries  should therefore
primarily depend on  the age of the system.   Fortunately, for a given
wide  WD/dM  binary  one can  estimate  the  age  of the  system,  and
consequently the  age of the M  dwarf companion, by  summing the white
dwarf    cooling    age    to    its    main    sequence    progenitor
lifetime\footnote{Note that  only reliable  ages can be  estimated for
  wide WD/dM binaries as PCEBs  have undergone a common envelope phase
  and an  initial-to-final mass  relation for the  white dwarf  is not
  valid  in these  cases.}.  SDSS  WD/dM binary  ages  were previously
calculated  by \citet{rebassa-mansergasetal12-1}  for a  sub-sample of
SDSS  WD/dM binaries. We  estimate the  ages for  the wide  SDSS WD/dM
binary sample studied  in this work (Section\,\ref{s-ident}) following
the same strategy.

The white  dwarf cooling ages  were calculated using  the evolutionary
tracks  of  of  \citet{bergeronetal95-2}\footnote{We  used  Bergeron's
  updated               grids               available               at
  http://www.astro.umontreal.ca/$\sim$bergeron/CoolingModels/}.     The
initial  main sequence  progenitor  masses were  calculated using  the
initial-to-final mass relation  by \citet{catalanetal08-1} and we used
the  equations  of   \citet[][see  their  Appendix]{tuffsetal04-1}  to
calculate their main sequence  lifetime.  In this exercise we excluded
wide WD/dM  binaries containing white dwarfs cooler  than 12000\,K, as
the  white  dwarf spectral  fitting  overestimates  the  mass in  such
systems   \citep[e.g.][]{koesteretal09-1,   tremblayetal11-1},   which
affects  the resulting  cooling  age.   For the  same  reason we  only
determined the age of wide WD/dM  binaries if the error of the mass of
the  white  dwarf is  smaller  than  $0.2\Msun$,  and given  that  the
initial-to-final mass  relation is only well defined  for systems with
white  dwarfs masses  $>$0.55$\Msun$,  we only  considered wide  WD/dM
binaries containing white dwarfs with masses above this value.

\begin{figure*}
\includegraphics[angle=-90,width=0.49\textwidth]{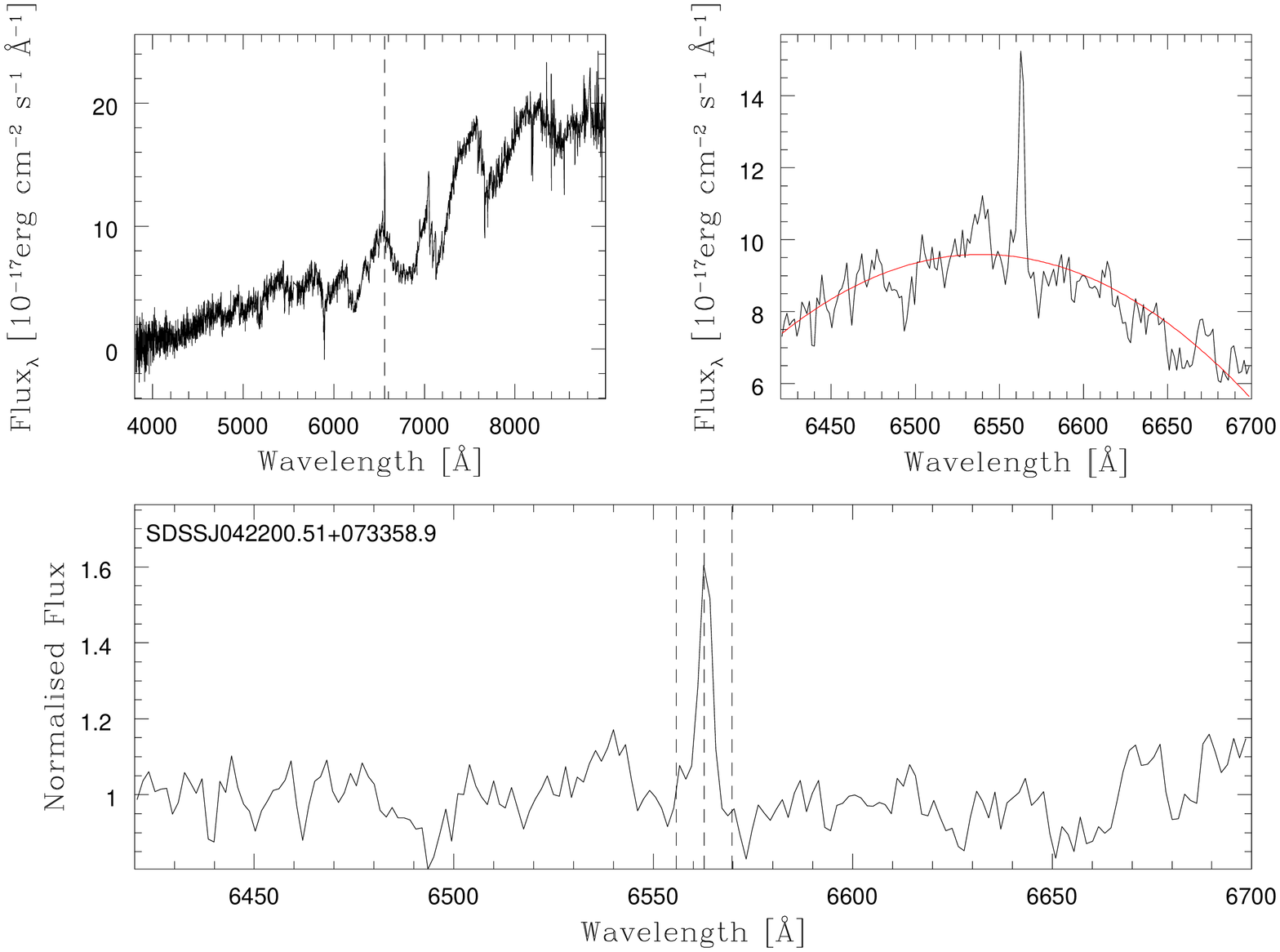}
\includegraphics[angle=-90,width=0.49\textwidth]{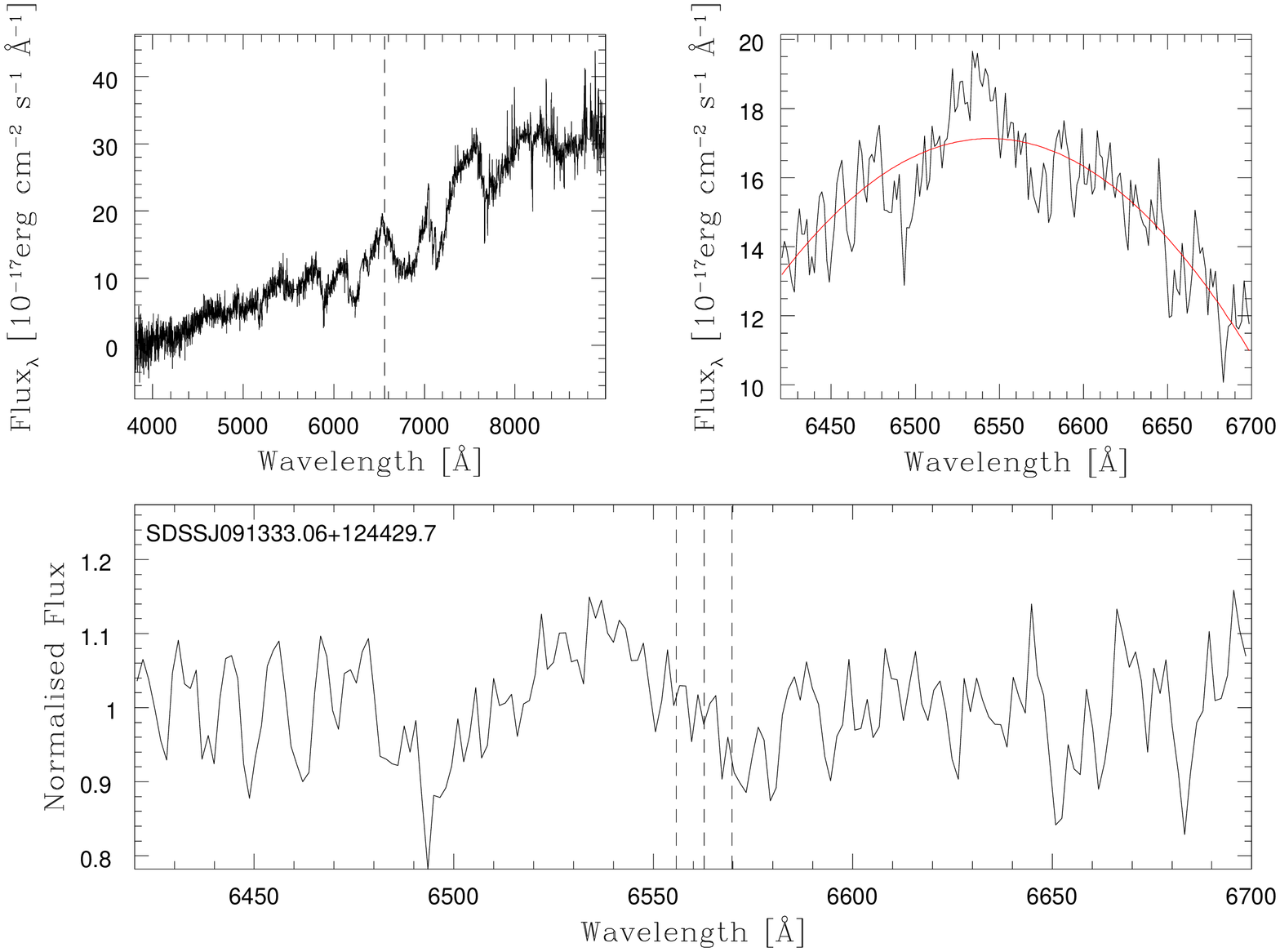}
\caption{\label{f-norm2}  Left:  the  residual  M  dwarf  spectrum  of
  SDSSJ042200.51+073358  (the   complete  spectrum  can   be  seen  in
  Figure\,\ref{f-norm1}) after  the subtraction of  the best-fit white
  dwarf model  (top left), the  parabolic fit to the  continuum around
  H$\alpha$ (red  solid line, top  right) and the  normalised spectrum
  (bottom).  The  vertical dashed lines indicate, from  left to right,
  Ha-7\AA, Ha  and Ha+7\AA\, where Ha=6562.76\,\AA\,  and Ha-7\AA\, $<
  \lambda <$ Ha+7\AA\, is the wavelength range in which the equivalent
  width of H$\alpha$ is  calculated.  In this case, H$\alpha$ emission
  is clearly detected,  and we catalogue the system  as active. Right:
  the same but for  SDSSJ091333.06+124429.7, an inactive system in our
  sample.}
\end{figure*}

\section{Identification of active M dwarfs}
\label{s-ident}

In the context of this paper it is necessary to unambiguously identify
both  active  and  inactive  M  dwarfs among  the  SDSS  WD/dM  binary
catalogue described  in Section\,\ref{s-cat}.  Two  features have been
predominantly  used as  activity indicators  in  optical spectroscopy,
namely  the  \Ion{Ca}{II}  H  and  K  resonance  lines  and  H$\alpha$
emission.   H$\alpha$ emission  is more  accessible in  WD/dM binaries
because the \Ion{Ca}{II}  H and K lines fall  at wavelengths generally
dominated by  the flux  of the white  dwarf.  We therefore  consider a
given M dwarf component as active if H$\alpha$ emission is detected in
its SDSS spectrum.  We developed an automatic routine for this purpose
that is described below.

In   a   first   step   the   best  white   dwarf   model   fit   (see
Section\,\ref{s-param}, Figure\,\ref{f-norm1})  is subtracted from the
SDSS  WD/dM binary  spectrum.  In  order  to avoid  our results  being
dominated by  low quality data, only  residual M dwarf  spectra with a
signal-to-noise    ratio   (S/N)    of   at    least   $>$    10   are
used\footnote{Originally a S/N of at  least $>$3 was used, however the
  activity  fractions   (see  details  in   Section\,\ref{s-frac})  we
  obtained  for  late-type  M  dwarfs  where  very  much  affected  by
  low-quality  data, and  we  therefore decided  to  increase the  S/N
  threshold  to a  least $>$10.},  which results  in a  sample  of 739
objects.  For these  739 systems the continuum flux  of the residual M
dwarf  around H$\alpha$ (6420\AA\,  $< \lambda  <$ 6700\AA)  is fitted
with a parabola (excluding the range Ha-7\AA\,$< \lambda <$ Ha$+7$\AA,
where Ha$=6562.76$\AA).   The parabolic fit is then  used to normalize
the spectrum.  Examples are  shown in Figure\,\ref{f-norm2}.  Once the
spectrum   is   normalized,   the   equivalent  width   of   H$\alpha$
(EW$_\mathrm{H\alpha}$)  is calculated within  the range  Ha-7\AA\,$ <
\lambda  <$  Ha$+7$\AA\, (emission  in  our  case  is indicated  by  a
negative  EW).   Active  M  dwarfs  are  then  selected  applying  the
following criteria:

\begin{eqnarray}
 \mathrm{EW}_\mathrm{H\alpha} \leq -0.75\mathrm{\AA}\\
 \mid \mathrm{EW}_\mathrm{H\alpha}\mid > 3 \times \mid \mathrm{eEW}_\mathrm{H\alpha}\mid\\
 \mathrm{h} > 3 \times \mathrm{N}_\mathrm{cont}
\end{eqnarray}

\noindent
where eEW$_\mathrm{H\alpha}$ is  the EW$_\mathrm{H\alpha}$ error, h is
the  peak of  the H$\alpha$  emission above  the continuum  level, and
N$_\mathrm{cont}$  is the  noise at  continuum level.   An M  dwarf is
considered inactive  if, conversely,  any of the  following conditions
apply:

\begin{eqnarray}
 \mathrm{EW}_\mathrm{H\alpha}>  -0.75\mathrm{\AA}\\
\mid \mathrm{EW}_\mathrm{H\alpha} \mid \leq 3 \times \mid \mathrm{eEW}_\mathrm{H\alpha}\mid\\
  \mathrm{h}   \leq   3   \times \mathrm{N}_\mathrm{cont}
\end{eqnarray}

\noindent

The above  procedure results in 373  active and 366  inactive M dwarfs
among the  739 SDSS  WD/dM binaries that  form our final  sample (note
that for  objects with multiple  SDSS exposures we consider  as active
those systems for which H$\alpha$ emission is detected in at least one
of the SDSS available M dwarf residual spectra).  Visual inspection of
the active and inactive samples confirms the efficiency of our method,
with  only three per  cent of  miss-classifications.  Note  that these
miss-classifications are due  to the fact that we  are subtracting the
white dwarf  contribution from the  SDSS spectra, which  may introduce
uncertainties in the  residual flux of the M dwarf  if the white dwarf
dominates the  spectral energy distribution  of the SDSS  WDMS binary.
Examples of active and inactive  M dwarf residual spectra are shown in
the bottom panels of Figure\,\ref{f-norm2}.

\begin{figure*}
\includegraphics[angle=-90,width=0.33\textwidth]{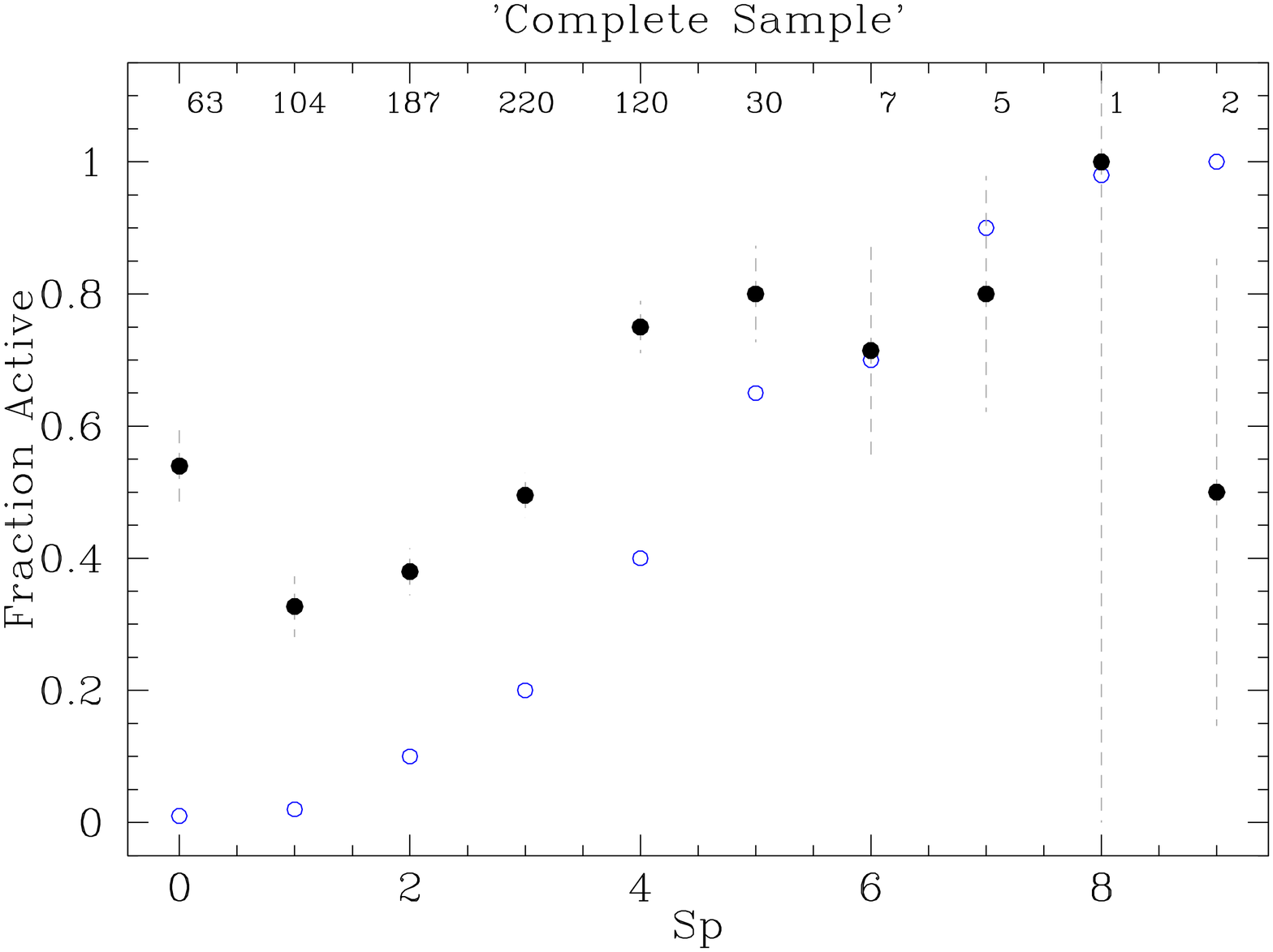}
\includegraphics[angle=-90,width=0.33\textwidth]{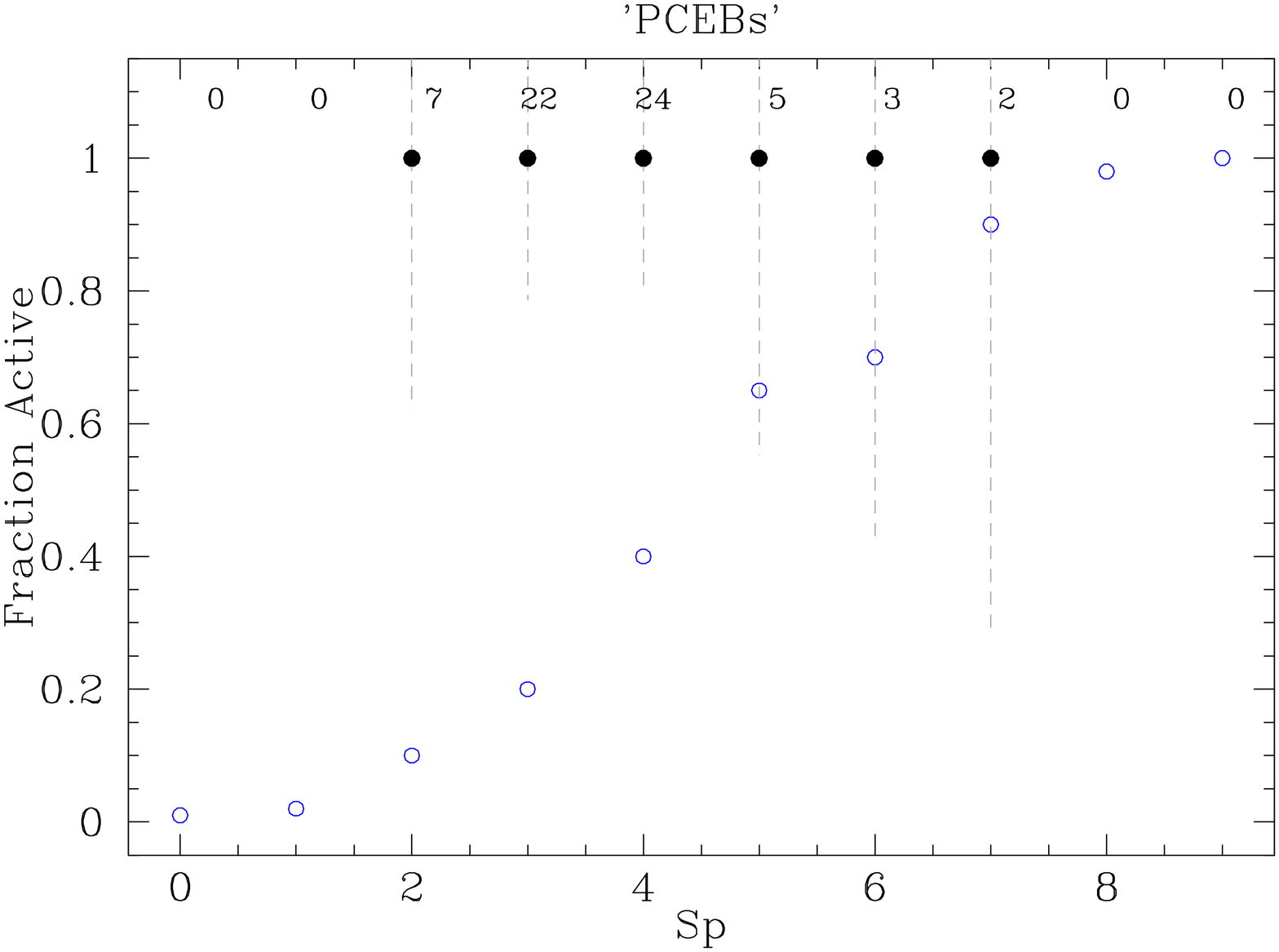}
\includegraphics[angle=-90,width=0.33\textwidth]{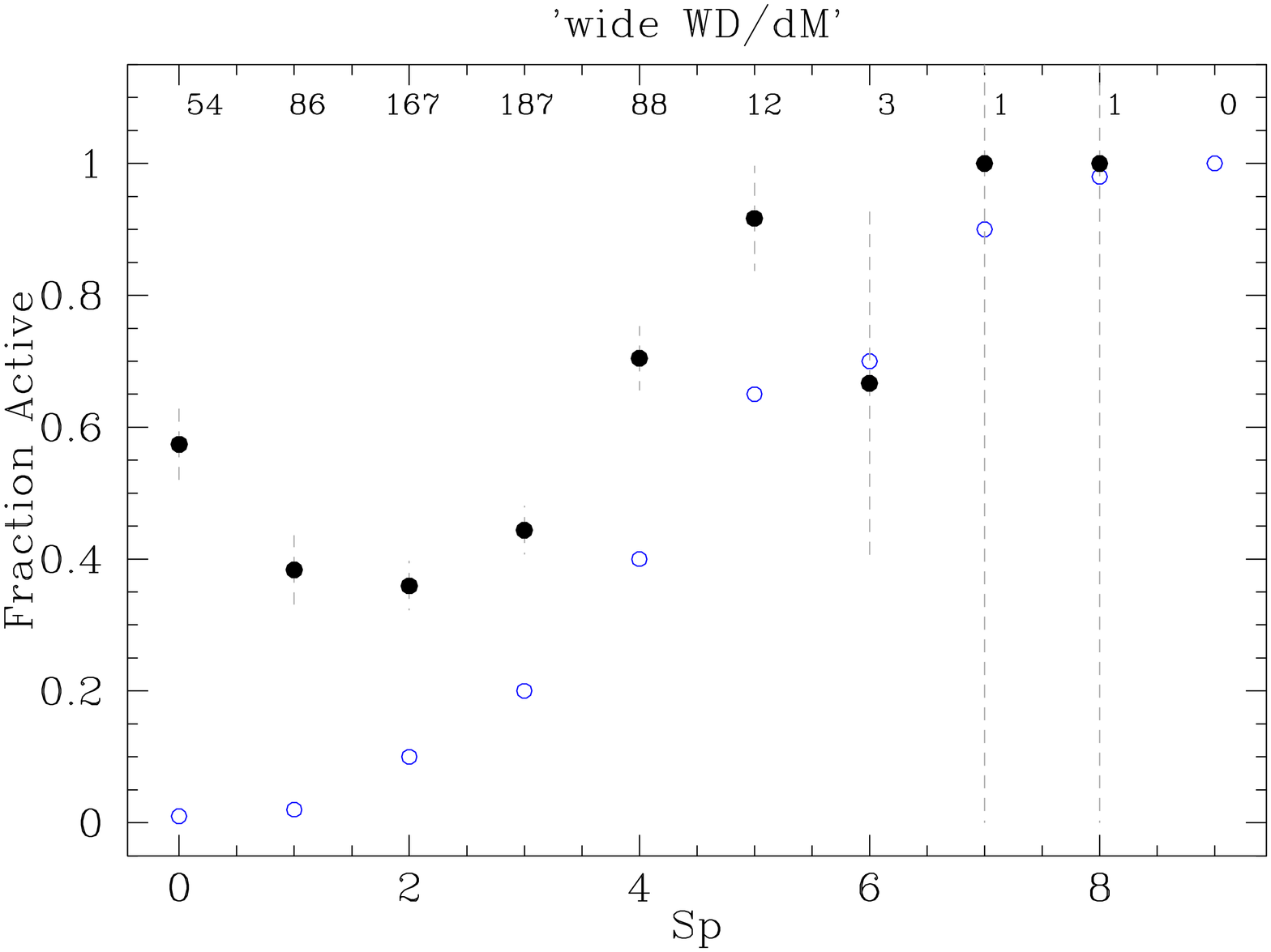}
\caption{\label{f-frac}  Activity fractions  (black solid  dots)  of M
  dwarfs in the complete SDSS WD/dM binary sample studied in this work
  (left),  in  close  PCEBs  (middle),  and  in  wide-separated  WD/dM
  binaries (right).   For comparison we  show the single SDSS  M dwarf
  activity fractions  (blue empty dots).  The total  number of systems
  per spectral type bin are given  at the top of each panel. Note that
  the total number of systems  on the top left panel (complete sample)
  are not the sum of the total number of systems on the top middle and
  right  panels (PCEB and  wide WD/dM  sub-samples respectively)  at a
  given spectral type because we have radial velocity information only
  for a fraction of the complete  sample.  Note also that the error in
  the activity fractions increases towards later spectral types due to
  the decrease in the number  of WD/dM binaries containing late type M
  dwarfs.}
\end{figure*}

From our radial  velocity survey (Section\,\ref{s-sample}) we identify
63 PCEBs and 599 wide WD/dM  binaries among the 739 systems.  Ages are
available for 118 of the 599 wide binaries (Section\,\ref{s-age}).

\section{Activity fractions}
\label{s-frac}

SDSS  single  M  dwarf  activity  fractions  have  been  published  by
\citet{westetal04-1,  westetal08-1, westetal11-1}  for  different SDSS
data  releases (DR\,4,  DR\,5  and DR\,7  respectively),  as well  for
independent (non-SDSS)  M dwarf samples \citep[e.g.][]{hawleyetal96-1,
  reinersetal12-1}.   All  show  the   same  general  trend,  i.e.   a
significant  increase of  the fraction  of active  systems  going from
partially to fully convective stars.

The M  dwarf activity fractions  of our 739  SDSS WD/dM binaries  as a
function  of   spectral  type   are  shown  on   the  left   panel  of
Figure\,\ref{f-frac} (black dots) together with the activity fractions
obtained by  \citet{westetal11-1} for single  M dwarfs from  the SDSS.
The  two  distributions differ  significantly  at early-to-mid  (M0-4)
spectral types.  At first glance this difference may not seem entirely
surprising, as  a significant fraction  of WD/dM binaries are  in fact
close PCEBs containing  rapidly rotating M dwarfs and,  as outlined in
the introduction, fast rotation  induces magnetic activity.  One might
therefore   speculate,    as   e.g.    \citet{silvestrietal06-1}   and
\citet{morganetal12-1}, that the  increased activity fraction of WD/dM
binaries containing early-to-mid spectral type M dwarfs is caused by a
fraction of short  orbital period PCEBs.  However, we  have shown that
very few  WD/dM binaries  containing early type  M dwarfs  ($<$M3) are
PCEBs  \citep{schreiberetal10-1}, i.e.   the majority  of  these PCEBs
have evolved  towards shorter orbital periods due  to angular momentum
loss via  magnetic braking  and entered a  semi-detached configuration
becoming cataclysmic variables \citep{politano+weiler06-1}.  Therefore
the  fraction of  PCEBs misclassified  as wide  systems in  our radial
velocity survey should be low in this spectral type range too.

Clearly, the  observed activity fraction  does not only depend  on the
rotational velocities but also on  the intrinsic ages of the M dwarfs.
Therefore, to  understand the measured activity  fractions requires to
separately  investigate close PCEBs  and wide  WD/dM binaries,  and to
take into account the ages of the latter population.

\subsection{PCEBs}
\label{s-pcebfrac}

The  SDSS PCEB orbital  period distribution  peaks at  $\sim$\,8 hours
\citep{nebotetal11-1},  and the  longest SDSS  PCEB orbital  period is
slightly below 10 days \citep{rebassa-mansergasetal12-2}. Consequently
M dwarfs in SDSS  PCEBs are tidally locked \citep{gladmanetal96-1} and
are expected to be magnetically active.

The  middle panel  of  Figure\,\ref{f-frac} shows  that  all M  dwarfs
within SDSS PCEBs are active  independently of the spectral type.  One
might be inclined  to interpret this result as  an obvious consequence
of rotation causing activity, nevertheless  it is important to keep in
mind  that H$\alpha$  emission  in PCEBs  may  arise from  irradiation
effects if the  white dwarf is relatively hot  and/or the PCEB orbital
period    is    relatively    short    \citep[e.g.][]{tappertetal11-1,
  tappertetal11-2}.   Therefore, we  investigate below  the  effect of
irradiation present among the 63 PCEBs in our sample.  For 47 of these
PCEBs the  orbital periods are  well known and the  stellar parameters
relatively   well   constrained   \citep{nebotetal11-1}.   With   this
information  at   hand  we  can  easily  estimate   the  influence  of
irradiation by calculating the ratio between the intrinsic flux of the
M  dwarf (F$_\mathrm{int}$) and  the irradiating  flux from  the white
dwarf on the M dwarf's surface (F$_\mathrm{irr}$),

\begin{equation}
\frac{F_\mathrm{int}}{F_\mathrm{irr}} = \left(\frac{T_\mathrm{dM}}{T_\mathrm{WD}}\right)^4 \times \left(\frac{a}{R_\mathrm{WD}}\right)^2,
\end{equation}

\noindent
where $T_\mathrm{dM}$  and $T_\mathrm{WD}$ are  the M dwarf  and white
dwarf  effective  temperatures  respectively, $R_\mathrm{WD}$  is  the
white dwarf radius, and $a$ is the orbital separation.

\begin{figure}
\includegraphics[width=\columnwidth]{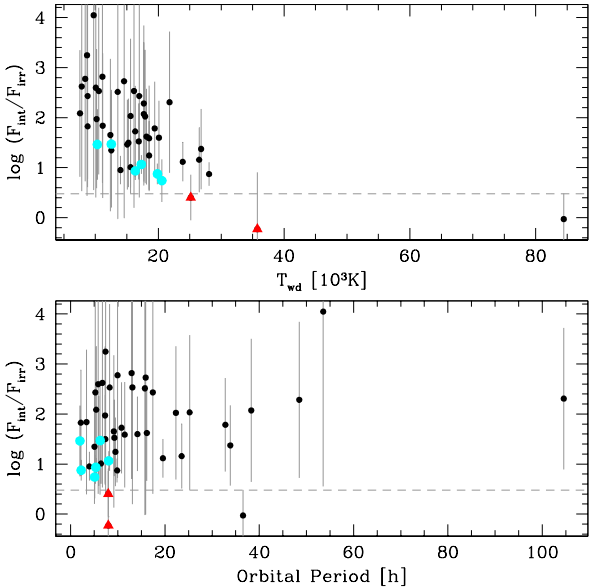}
\caption{\label{f-ratio}  The M  dwarf  intrinsic flux  / white  dwarf
  irradiated  flux   ratio  (F$_\mathrm{int}$/F$_\mathrm{irr}$)  as  a
  function  of white  dwarf  effective temperature  (top) and  orbital
  period  (bottom).   The   horizontal  gray  dashed  line  represents
  F$_\mathrm{int} =\,  3\times$ F$_\mathrm{irr}$ and is  adopted to be
  the maximum  limit for irradiation  effects to be strong  enough for
  H$\alpha$  to  be  in   emission.   Red  solid  triangles  represent
  confirmed PCEBs showing signs  of irradiation effects in their light
  curves, cyan solid dots PCEBs  displaying no signs of irradiation in
  their light  curves.  The large  errors obtained for several  of our
  systems  are a  simple consequence  of propagating  relatively large
  uncertainties    on    the    estimated    effective    temperatures
  \citep{rebassa-mansergasetal07-1}.}
\end{figure}

The  resulting F$_\mathrm{int}$/F$_\mathrm{irr}$  ratios are  shown in
Figure\,\ref{f-ratio}   as  a  function   of  white   dwarf  effective
temperature (top panel) and  orbital period (bottom panel).  Among the
47 PCEBs, eight were followed-up photometrically as part of our survey
(Section\,\ref{s-sample}) and inspection of their light curves reveals
clear irradiation effects in only  two objects (red solid triangles in
Figure\,\ref{f-ratio}, cyan dots represent the remaining six systems).
Based  on  the  evidence  drawn  from  these  eight  PCEBs  we  define
F$_\mathrm{int}  =\,   3\times$  F$_\mathrm{irr}$  as   the  threshold
separating  systems  dominated  by  irradiation from  those  that  are
dominated by  the intrinsic emission  of the M dwarf  (horizontal gray
dashed line in Figure\,\ref{f-ratio}).  93 per cent of our systems lie
above  this  limit  and  are  presumably  unaffected  by  irradiation.
Although the limit defined in  that way just represents an empirically
motivated  estimate, we  conclude that  irradiation  induced H$\alpha$
plays a minor role for our sample of PCEBs and that \emph{all M dwarfs
  in PCEBs are magnetically active}.

\subsection{Wide WD/dM binaries}
\label{s-wide}

It is  generally accepted  that the stellar  components in  wide WD/dM
binaries have  evolved in the  same way as  if they were  single stars
\citep{willems+kolb04-1}.  Consequently the  properties of the M dwarf
companions  in such  binaries should  be  fairly similar  to those  of
single  M  dwarfs,  and  naively  one would  expect  similar  activity
fractions in both populations.  This  is clearly not the case (see the
right panel of Figure\,\ref{f-frac}).

The larger activity fractions detected in our sample must be caused by
the fact that these M dwarfs  are part of binary systems, but as these
binaries                 are                fairly                wide
\citep[$\Porb>100$\,days,][]{willems+kolb04-1},  interactions  between
the two  stars can  be excluded to  to be  the cause of  the increased
activity  fractions.   As  mentioned in  Section\,\ref{s-sample},  our
radial velocity survey suffers from observational biases and we expect
a small  percentage of the  wide WD/dM binary  sample to be  formed by
unrecognised  PCEBs.  It  is therefore  important to  evaluate whether
this effect can explain the increase of the activity fractions seen in
M dwarfs that are part of  wide WD/dM binaries as compared to those of
single M  dwarfs. For  this purpose we  calculate in what  follows the
PCEB contamination in our wide WD/dM binary sample at a given spectral
type.  The exact number of  unrecognised PCEBs depends on the spectral
type   of   the   M   dwarf   components.    \citet{schreiberetal10-1}
demonstrated that the observed fraction of PCEBs containing early-type
M dwarfs is $\sim$10 per cent at M0-1 spectral types, and increases to
only $\sim$20 per cent at M2.  Combining these fractions with the PCEB
detection  probability  of our  radial  velocity  survey of  $\sim$0.8
\citep{schreiberetal10-1,  rebassa-mansergasetal11-1}  we calculate  a
PCEB contamination of less than five per cent in the M0-2 range (where
the  activity  fractions  most  differ).  Therefore  this  effect  can
certainly not explain the difference between the activity fractions in
the   single  and   wide  WD/dM   binary  samples   (right   panel  of
Figure\,\ref{f-frac}).

Identifying both  components in a SDSS WD/dM  binary spectrum requires
that neither the white dwarf  nor the M dwarf completely outshines the
other component.   Cold white dwarfs are  too faint to  be detected in
binaries with  early-type M dwarfs.   This implies the white  dwarf to
have  an  effective temperature  $\gappr10000$\,K  thus excluding  the
coolest (i.e.   oldest) white dwarfs \citep{rebassa-mansergasetal07-1,
  rebassa-mansergasetal10-1}.\footnote{Note  that the decrease  of the
  number  of  PCEBs containing  early  type M  dwarfs  is  due to  two
  separate reasons. First, the  majority of these systems have already
  entered  a   semi-detached  configuration  and   thus  evolved  into
  cataclysmic  variables   \citep{schreiberetal10-1}.   Second,  white
  dwarf companions to early type M dwarfs need to be relatively hot to
  be detected in  the SDSS spectra, therefore the  S/N of the residual
  SDSS M dwarf spectra in these objects is generally lower than 10 and
  are excluded from  the analysis (Section\,\ref{s-ident}).  These two
  effects result  in no  systems with M0--1  M dwarf  components being
  included    in   the    PCEB   sample    studied   in    this   work
  (Figure\,\ref{f-frac},  middle panel).}   The  mass distribution  of
white  dwarfs  in   wide  WD/dM  binaries  is  similar   to  the  mass
distribution of  single white dwarfs \citep{rebassa-mansergasetal11-1}
implying that the progenitors of such white dwarfs were mainly F stars
that spent a rather short time  on the main sequence.  In other words,
the simple  requirement that we  need to be  able to detect  the white
dwarf  in the  SDSS  spectrum to  classify  a system  as WD/dM  binary
introduces  a  significant bias  towards  young  systems  that is  not
present in the samples of single M dwarfs.

\begin{figure}
\centering
\includegraphics[angle=-90,width=\columnwidth]{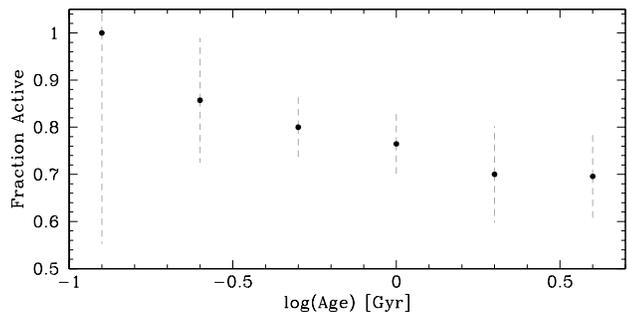}
\caption{\label{f-age} Fraction  of active M dwarfs that  form part of
  wide WD/dM  binaries as a function  of age. For clarity,  we bin the
  data by age.  The data suggests a decrease in activity fraction with
  increasing M-dwarf age, in very good agreement with rotation braking
  with time.}
\end{figure}

In  order   to  investigate  this   hypothesis  further  we   show  in
Figure\,\ref{f-age} the  activity fractions as  a function of  age for
118    wide   WD/dM   binaries    with   available    estimated   ages
(Section\,\ref{s-age}).   Two  important  results are  extracted  from
inspecting  this figure.   (1) As  expected,  M dwarfs  in wide  WD/dM
binaries are in general significantly  younger than the average age of
field  M dwarfs  of $\sim5\,$Gyrs  \citep{westetal08-1}.   On average,
younger  stars  rotate  faster  and therefore  a  considerably  higher
fraction of  M dwarfs in  WD/dM binaries are  active.  (2) There  is a
clear  decrease of  the activity  fractions for  older  systems.  Even
though the errors in the  activity fractions are fairly broad, this is
in very good agreement with magnetic braking slowing down the rotation
of the M dwarfs that are part of wide WD/dM binaries with time.

It is  worth mentioning that  the activity fractions measured  for the
systems with available ages seem  to be higher than those obtained for
the  complete wide  WD/dM binary  sample (see  Figure\,\ref{f-age} and
right panel  of Figure\,\ref{f-frac}).  This is simply  because we can
only  obtain reliable  ages  if the  white  dwarfs are  hot enough  to
dominate the SDSS WD/dM binary spectra, which reflects in an effective
temperature of $\gappr15000-20000$\,K,  depending on the spectral type
of the M dwarf.  Thus  an older population containing inactive systems
exists in  our wide WD/dM binary sample  for which we are  not able to
estimate  the  ages,  making  the  activity fractions  lower  for  the
complete wide  WD/dM binary sample but still  considerably higher than
those measured for field M dwarfs.

We  conclude that  the  age  effect described  above  can explain  the
observed  difference  between   the  activity  fractions  measured  by
\citet{westetal11-1} for  single M dwarfs and those  obtained here for
the M dwarfs in wide WD/dM binaries.

\section{The rotation-activity relation}
\label{s-rot}

As outlined  in the introduction,  rotation and magnetic  activity are
strongly correlated. The strength  of activity increases with rotation
and  saturates around  $R_{0}  \simeq  0.1$, the  exact  value of  the
saturation  threshold  velocity depending  on  the  mass  of the  star
\citep{pizzolatoetal03-1}.   Such a  saturation-type rotation-activity
relation has  been observed  for the entire  spectral type range  of M
dwarfs  \citep{mohanty+basri03-1, reiners+basri10-1, reinersetal12-1}.

In  this Section we  investigate the  saturation-type relation  in our
sample  of   47  PCEBs   for  which  we   know  the   orbital  periods
(Section\,\ref{s-pcebfrac}).   For  this   purpose,  the  strength  of
magnetic activity of the 47 systems is quantified by the ratio between
the        H$\alpha$        and        bolometric        luminosities,
L$_\mathrm{H\alpha}$/L$_\mathrm{bol}$.    This  luminosity   ratio  is
obtained   from  the   measured   EW$_\mathrm{H\alpha}$  values   (see
Section\,\ref{s-ident})      following      the      equations      of
\citet{west+hawley08-1}, who  obtained conversion factors  between the
two  considered quantities. Compared  to single  field stars,  M dwarf
companions in PCEBs  are tidally locked and are  therefore expected to
be very fast rotators that should populate the saturated regime.

\begin{figure}
\includegraphics[width=\columnwidth]{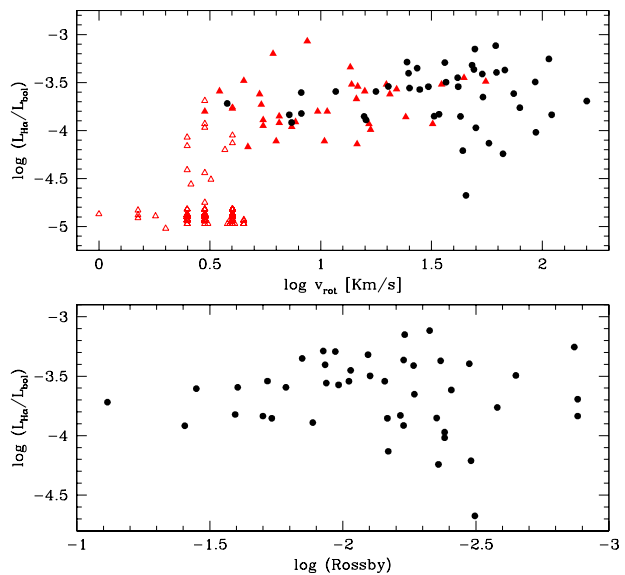}
\caption{\label{f-vrot} Top  panel: the strength  of magnetic activity
  (evaluated    through    the   L$_\mathrm{H\alpha}$/L$_\mathrm{bol}$
  luminosity   ratio)   as   a   function   of   rotational   velocity
  (V$_\mathrm{rot}$) of 47  M dwarfs in close PCEBs.   Our data (black
  solid     dots)     extend      the     work     by     \citet[][red
    triangles]{reinersetal12-1}  to faster rotators  and unambiguously
  confirm the saturation limit of the strength of magnetic activity at
  $\log$\,L$_\mathrm{H\alpha}$/L$_\mathrm{bol}   \sim  -3.5$.   Bottom
  panel: the strength of magnetic activity as a function of the Rossby
  number.           The          increased         scatter          of
  $\log$\,L$_\mathrm{H\alpha}$/L$_\mathrm{bol}$   at  $\log\,(Ro)  \la
  -1.8$ may indicate chromospheric supersaturation}
\end{figure}

Figure\,\ref{f-vrot}  (top panel)  shows  the activity  strength as  a
function of  rotational velocity V$_\mathrm{rot}$  for the 47  M dwarf
companions in PCEBs with well known orbital periods (black solid dots)
and  the   single  M  dwarfs   from  the  catalogue   of  \citet[][red
  triangles]{reinersetal12-1}.  Note that the latter sample is divided
into systems with accurately measured rotational velocities (solid red
triangles) and systems with  upper limits to the rotational velocities
(open  red  triangles),  and  that  these values,  contrary  to  ours,
represent  projected  rotational  velocities  (V$_\mathrm{rot}  \times
\sin\,i$).   Inspection  of  the  top  panel  of  Figure\,\ref{f-vrot}
clearly reveals that the M dwarfs  in PCEBs populate the regime of the
fastest rotators in which  the strength of magnetic activity saturates
at  $\log$\,L$_\mathrm{H\alpha}$/L$_\mathrm{bol}   \sim  -3.5$.   This
result  represents  a crucial  and  very  robust  confirmation of  the
proposed  rotation-activity   relations  in  M   dwarfs.   To  further
illustrate   this   point,   we   show   in  the   bottom   panel   of
Figure\,\ref{f-vrot} the activity strength as a function of the Rossby
number  $R_{0}$.   The latter  has  been  calculated  as described  in
\citet{reiners+basri10-1}.  All  our systems fall  below the threshold
of $R_{0} \simeq 0.1$, well within the saturated regime.

A  closer  inspection  of  the bottom  panel  of  Figure\,\ref{f-vrot}
reveals that whilst practically all systems with $\log\,(R_\mathrm{o})
\gappr -1.8$ have -3.5$\la \log$\,L$_\mathrm{H\alpha}$/L$_\mathrm{bol}
\la -4.0$,  there is substantial system-to-system  scatter for smaller
Rossby  numbers.    Interestingly,  \citet{staufferetal97-1}  found  a
decline in  X-ray saturation for  stars with $R_\mathrm{o} <  0.01$ in
the  coronal  emission of  young  G and  K  dwarfs,  an effect  termed
``supersaturation''.  Whilst  evidence for coronal  supersaturation in
F,G,K    dwarfs    is    growing    \citep{wrightetal11-1},    coronal
supersaturation    has    not   been    identified    in   M    dwarfs
\citep{jamesetal00-1, jeffriesetal11-1} and evidence for chromospheric
supersaturation has not been identified  neither in F,G,K nor M dwarfs
\citep{marsdenetal09-1, jackson+jeffries10-1}.  While  our data do not
show a  statistically significant decline of the  strength of activity
for  very  fast  rotators  (top panel  of  Figure\,\ref{f-vrot}),  the
increased scatter of the chromospheric activity level for systems with
very small  Rossby numbers  in our PCEBs  sample is intriguing  in the
context  of supersaturation.  One  could speculate  that we  see first
indications  of chromospheric supersaturation  in M  dwarfs.  However,
the   current  data   are  clearly   not  conclusive   and  additional
measurements of  chromospheric activity of  M dwarfs in  fast rotating
PCEBs are required to test this hypothesis.

\section{Summary and conclusions}
\label{s-disc}

Angular momentum  evolution, rotation and  age are closely  related in
low-mass  stars. Magnetic  field  generation strongly  depends on  the
Rossby   number,   therefore   the   rotation,   of   a   given   star
\citep{hartmann+noyes87-1,   chabrier+kueker06-1}.    Thus,   magnetic
activity increases with rotation until its strength saturates for fast
rotation  rates \citep{pizzolatoetal03-1,  reinersetal09-1}.  Magnetic
braking slows down the rotation  of the star and consequently magnetic
activity   and  rotation   decrease   in  time   \citep{skumanich72-1,
  browningetal10-1}.  The timescale of magnetic braking depends on the
mass of  the star,  i.e.  in fully  convective stars  magnetic braking
becomes  inefficient \citep{reiners+basri08-1,  schreiberetal10-1} and
the activity lifetimes increase \citep{westetal08-1}.

WD/dM  binaries  offer   an  excellent  opportunity  to  independently
investigate the relations among magnetic activity, rotation and age in
M dwarfs.  This  is because WD/dM binaries can  be separated into wide
binaries that never  interacted, and in which the  properties of the M
dwarf components should be fairly similar to those of single stars, as
well  as  close  binaries  that underwent  common  envelope  evolution
(PCEBs)  in which  the  M dwarf  is  tidally locked  and rotates  very
rapidly      \citep{rebassa-mansergasetal11-1,      schreiberetal10-1,
  nebotetal11-1}.  To  test the activity-rotation-age  connection with
the largest and  most homogeneous sample of SDSS  WD/dM dwarf binaries
currently            known           \citep{rebassa-mansergasetal10-1,
  rebassa-mansergasetal12-1} has been the main goal of this paper. Our
main results can be summarised as follows:

\begin{itemize}

\item
We have demonstrated that for  M dwarfs in PCEBs fast rotation implies
activity  independently  of  the  spectral type.   This  supports  the
connection between rotation and activity as a necessary ingredient for
dynamo  models  responsible  for  magnetic field  generation  in  both
partially    and   fully    convective    stars   \citep{browning08-1,
  chabrier+kueker06-1}.

\item
We  have shown  that M  dwarfs in  close WD/dM  binaries  populate the
saturated  regime ($Ro  <  0.1$)  and that  the  strength of  magnetic
activity   saturates  at  $\log$\,L$_\mathrm{H\alpha}$/L$_\mathrm{bol}
\sim   -3.5$.   This   represents  a   crucial  confirmation   of  the
saturation-type rotation-activity relation for  M dwarfs that are very
fast   rotators,   in   perfect   agreement  with   the   results   by
\citet{reiners+basri10-1, reinersetal12-1} for single M dwarfs.

\item
We  have  provided  additional  observational  evidence  for  magnetic
braking  becoming  inefficient in  fully  convective  stars i.e.   the
activity  fractions  of  M  dwarfs  in  wide  WD/dM  binaries  show  a
significant increase  at the fully convective boundary.   We have also
shown that the  activity fractions of M dwarfs  in wide WD/dM binaries
decrease in time, which proves the  spin-down of these M dwarfs due to
magnetic braking.

\end{itemize}

We  conclude the  observational findings  from the  SDSS  WD/dM binary
sample  are consistent with  the current  picture of  angular momentum
evolution and dynamo generation  of magnetic fields in low-mass stars,
and add  important information to  our understanding of  the relations
between  magnetic activity, rotation,  age and  magnetic braking  in M
stars.

\section*{Acknowledgments.}

ARM  acknowledges  financial   support  from  Fondecyt,  grant  number
3110049.  MRS  acknowledges support from  Milenium Science Initiative,
Chilean   Ministry  of  Economy,   Nucleus  P10-022-F,   and  Fondecyt
(1100782). We thank Ada  Nebot Gomez-Moran for helpful discussions and
the anonymous  referee for  his/her suggestions that  helped improving
the quality of the paper.

\label{lastpage}

\end{document}